\documentclass{llncs}
\usepackage{amsmath}
\usepackage{graphicx}
\usepackage{amssymb,euscript}
\usepackage[english]{babel}
\usepackage{epstopdf}

\begin{document}

\title{NUMERICAL SOLUTION OF A PARABOLIC SYSTEM IN AIR POLLUTION}

\author{Tatiana P. Chernogorova,  Lubin G. Vulkov}
\institute{FMI, Sofia University, 5, J. Bourchier Blvd., 1164 Sofia, Bulgaria,\\
FNSE, Uiversity of Rousse, 8, Studentska str., 7017 Rousse, Bulgaria\\
e-mails:\{chernogorova@fmi.uni-sofia.bg, lvalkov@uni-ruse.bg \}}

\maketitle

\begin{abstract}
An air pollution model is generally described by a system of PDEs on unbounded domain.
Transformation of the independent variable is used to convert the problem for nonlinear air pollution
on finite computational domain. We investigate the new, degenerated parabolic problem in
Sobolev spaces with weights for well-posedness and positivity of the solution. Then we
construct a fitted finite volume difference scheme. Some results from computations are presented.
\end{abstract}



\section{Introduction}

Environmental problems are becoming more and more important for our world
and their importance will even increase in the future. High pollution of
air, water and soil may cause damage of plants, animals and humans.

An air pollution model is generally described by a system of PDE-s for
calculating the concentrations of a number of chemical species (pollutants
and components of the air that interact with the pollutant) in a large
3-D domain (part of the atmosphere above the studied geographical region).
The Danish Euler Model (DEM) is one of the most frequently used air pollution model and
is mathematically represented by the following system PDE-s
{\cite{13}}, {\cite{10}}, {\cite{8}}, {\cite{11}}:
\begin{multline*}
\frac{\partial c_s }{\partial t} = - \frac{\partial (uc_s )}{\partial x} -
\frac{\partial (vc_s )}{\partial y} - \frac{\partial (wc_s )}{\partial z} +
\frac{\partial }{\partial x}\left( {K^{x}_{s} \frac{\partial c_s }{\partial x}}
\right) + \frac{\partial }{\partial y}\left( {K^{y}_{s} \frac{\partial c_s}{\partial y}} \right) \nonumber \\
+ \frac{\partial }{\partial z}\left( {K_{s}^{z}\frac{\partial c_s }{\partial z}} \right)
 + F_s + R_s (c_1,c_2 ,\ldots ,c_S ) - (k_{1s} + k_{2s} )c_s ,\quad s =1,2,\ldots,S,
\end {multline*}
where $c_s$ are the \emph{concentrations} of the chemical species; $u$, $v$ and  $w$ are
wind velocities and $K_{s}^{x}$, $K_{s}^{y}$, $K_{s}^{z}$ are the
diffusion components; $F_s$ are the emissions; $k_{1s}$, $k_{2s}$ are
dry/wet deposition coefficients and $R_s(c_1, c_2,\ldots, c_S )$ are
non-linear functions describing the chemical reactions between the species under
consideration {\cite{11}}. Typical is the case
$R_{s}(c_{1}, c_2,\dots, c_{s})=\sum_{i=1}^{S}\gamma_{s,i} c_{i} +
\sum_{i=1}^{S}\sum_{j=1}^{S}\beta _{s,i,j} c_{i}c_{j}$, $s =1, 2, \dots, S,$
where $\gamma_{s,i}$ and $\beta_{s,i,j}$ are constants.

For such complex models operator splitting is very often applied in order to
achieve sufficient accuracy as well as efficiency of the numerical
solution. Although the splitting is a crucial step in efficient numerical
treatment of the model, after discretization of the large computational
domain each sub-problem becomes itself a huge computational task. Here we
will concentrate on a non-stationary sub-model of a horizontal advection-diffusion with chemistry,
emissions and deposition, see {\cite{8}}, {\cite{11}}:
\begin{gather}
\frac{\partial c_s }{\partial t}
-\frac{\partial }{\partial z}\left( {K_s (z)\frac{\partial c_s }{\partial z}}
\right)+ w\frac{\partial c_s }{\partial z}
- R_s (c_1, c_2,\ldots, c_S ) = Q_s (t) \delta (z - z_s^\ast ),\label{1}\\
 z \in (0,\infty),\quad t \in (0,T],\nonumber\\
\frac{\partial c_s }{\partial z}(t,0) = \delta _s c_s (t,0), \quad \delta _s = const \ge 0,
\quad t \in [0,T],\label{2}\\
\mathop {\lim }\limits_{z \to \infty } c_s (t,z) = 0, \quad t \in [0,T] \label{3},\\
c_s(0,z)=c_{s,0},\quad z \in [0,\infty),  \quad s = 1,2,\ldots,S. \label{4}
\end{gather}
The rest of the paper is organized  as follows. In Section 2 we investigate the transformed
differential problem and discuss its well-posedness  and the properties of its solution. In Section 3 we derive the fitted finite volume  discretization. In Section 4 we present some results from computational experiments. At the end we formulate some conclusions.
Let us mention that construction and analysis of positive numerical methods is very important for models in medicine and finance, see e.g.\cite{K13,K15}.
\section {The differential problem on bounded domain}

In the numerical scheme it is not convenient to corporate the boundary
condition at infinity. For the simplest case of (\ref{1})
(one linear advection-diffusion equation)
(discrete) transparent boundary conditions
are constructed and analyzed in {\cite{3}} while in {\cite{4}} the transformation
\[
z = \frac{1}{2a}\log \left( {\frac{1 + \xi }{1 - \xi }} \right),\quad
\xi \in \Omega = (0,1) \Leftrightarrow \xi = \frac{e^{2az} - 1}{e^{2az} +1},\quad z \in (0,\infty ),
\]
is used. Here $a$ is a stretching factor.
Using this transformation, the system (\ref{1}) and
the respective boundary and initial conditions (\ref{2})--(\ref{4})
in the computational domain become
\begin{gather}
\frac{\partial C_s }{\partial t} - a^2\left( {1 - \xi ^2} \right)^2k_s (\xi)
\frac{\partial ^2C_s }{\partial \xi ^2}  \nonumber
+ a\left( {1 - \xi ^2}\right)\left( {  2a\xi k_s (\xi ) + w - a\left( {1 - \xi ^2}
\right)\frac{\partial k_s (\xi )}{\partial \xi }} \right)\\ \times \frac{\partial C_s}{\partial \xi }
 - r_s (C_1,C_2 ,\ldots ,C_S ) = Q_s (t)\delta \left( {\xi - \xi_s ^\ast }
\right),\quad \xi \in \Omega, \; t \in (0,T], \label{5} \\
a\frac{\partial C_{s}}{\partial \xi}(t,0)= \delta_{s}C_{s}(t,0), \quad t \in [0,T],\label{6}\\
C_{s}(t,1) = 0, \quad t \in [0,T], \label{7}\\
C_s(0,\xi)=C_{s,0},\quad \xi \in [0,1],  \quad s = 1,2,\ldots,S.\label{8}
\end{gather}
We have denoted
$C_{s}(t,\xi) \equiv c_{s}(t,z(\xi))$, $k_s(\xi)=K_{s}(z(\xi))$, $r_{s}(C_{1},C_{2},\dots,C_{s})=
R_{s}(c_{1}(t,z(\xi)), c_{2}(t,z(\xi)),\dots,c_{S}(t,z(\xi))$.

It is easy to seen that at $\xi=1$ the system (\ref{5}) degenerates to the ODE system
\begin{equation}\label{9}
\frac{\partial C_{s} (t,1)}{\partial z}-r_{s}
(C_{1} (t,1), \dots, C_{S}(t,1)=0, \quad s = 1, 2, \dots, S, \quad t \in (0,T].
\end{equation}
In the case $C_{s,0}=0$, the unique solution of (\ref{9}) is the zero one, i.~e. (\ref{7}).

By the  Fichera and Oleinik-Radkevich theory {\cite{5}} for degenerate parabolic
equations, at the degenerate boundary $\xi =1$, the boundary condition should
not be given. But from physical motivation we have imposed the boundary
condition (\ref{7}). It is easy to check that if the functions
$C_{s}(t,\xi)$ satisfy (\ref{7}) then they also satisfy (\ref{9}).
Therefore (\ref{7}) is a particular case of (\ref{9}).

The general theory {\cite{5}} does not provide existence and uniqueness of the solution to the problem (\ref{5})--(\ref{8}).
Also, following the physical motivation we will discuss the non-negativity of the solution.
For this we rewrite the system (\ref{5}) in divergent form:
\begin{gather}
\frac{\partial C_s }{\partial t} = \frac{\partial }{\partial \xi }\left({p_s (\xi )\frac{\partial C_s }
{\partial \xi } + q_s (\xi )C_s } \right) +B_s (\xi, C_1,\ldots ,C_S ) +
f_s(t,\xi),\label{10} \\
\quad (t,\xi) \in  Q_{T} = [0,T] \times \Omega, \quad p_s (\xi ) = a^{2} ( 1 - \xi^{2} ) k_{s} ( \xi ),\nonumber \\
q_s(\xi )= a(1-\xi^{2})(2a \xi k_{s}(\xi)- w),\;
B_s (\xi,C_1,\ldots,C_S ) =r_{s}(C_{1}, \dots, C_{S})-d_s(\xi)C_s,\nonumber\\
d_s(\xi) = 2 a^{2}(1- 3\xi^{2} )k_{s}(\xi)+2a^{2}\xi(1-\xi^{2})
\frac{\partial k_{s}(\xi)}{\partial \xi}+2aw \xi,\; s = 1,2,\ldots, S,\nonumber
\end{gather}
where $f_s(t,\xi)$ is a regularization of the Dirac delta-function, $s = 1,2,\ldots, S$.

Further, to handle the degeneracy in the equation (\ref{10}), we introduce the weighted inner product
and corresponding norm on $L_{2,w} (\Omega )$ by
\[
(u,v)_{w} :=\int_{0}^{1}(1- \xi)^{2} u v d \xi, \quad
\Vert v \Vert_{0,w}={\sqrt {(v,v)_{w}}}=\left(\int_{0}^{1}(1- \xi)^{2} v^{2} d \xi\right)^{ 1/2}.
\]
The space of all weighted square-integrable functions is defined as
$L_{2,w} ( \Omega ) : =\{ v: \Vert v \Vert_{0,w} < \infty \} $.  By using a standard
argument it is easy to show that the pair $( L_{2,w}(\Omega),(\cdot, \cdot)_{w})$
is a Hilbert space (cf., for example, {\cite {6}}). Using $L_{2}(\Omega)$ and $L_{2,w}(\Omega)$, we define the following weighted Sobolev space
\[
H_{w}^{1}(\Omega):=\{v \in L_{2}(\Omega), \; v' \in L_{2,w}(\Omega) \},\quad v'=\partial v / \partial \xi
\]
with corresponding inner product on
$H_{w}^{1}(\Omega):=(\cdot, \cdot)_{H_{w}^{1}}:=(\cdot, \cdot)_{w}+(\cdot, \cdot)$.
Also, it is  easy  to prove that the pair
$ ( H_{w}^{1}  ( \Omega ), ( \cdot , \cdot )_{H_{w}^{1}  } )$ is a Hilbert space with the norm
\[
\Vert v \Vert_{1,w}:=\{\Vert v' \Vert_{0,w}^{2} +\Vert v \Vert_{2}^{2} \}^{1/2}=
\{ \left((1- \xi)^{2} v', v'\right) +(v,v) \}^{ 1/2}.
\]
\noindent
For $C_{s}, \eta_{s} \in H^{1}_w (\Omega )$ we define the bilinear forms
\[
A_{s} (C_s, \eta_s ;t) = \int_{0}^{1}\left[p_{s} ( \xi )\frac{ \partial C_{s} } {\partial \xi }
\frac{ \partial \eta_{s} } {\partial \xi } +q_{s} ( \xi ) C_{s}\frac{ \partial \eta_{s} } {\partial \xi }
-B_{s} ( \xi ,C_1, \ldots, C_S)C_s \eta_{s}
\right]d \xi.
\]
Now we are in position to define the following variational
problem corresponding to (\ref{10}) and (\ref{6}), (\ref{8}), (\ref{9}):
find $C=(C_1, \ldots, C_S)$, $C_{s} \in C ( [0,T]; H_{w}^{1} ( \Omega ))$, $s=1,2,\ldots,S$, satisfying the initial condition (\ref{6}), such that for all
$\eta_{s} \in H_{w}^{1} ( \Omega )$
\[
\int_{0}^{1}\frac{ \partial C_{s} } {\partial t } \eta_{s} d \xi + A_{s} (C_s, \eta_s;t) =
\int_{0}^{1} f_{s} \eta_{s} d \xi \; \; \mbox{a.~e in } (0,T).
\]
{\bf{Theorem 1.}}
Let $B_{s} (\xi, C) $ be Lipshitz continuous with respect to $C$. There exists a unique solution $C$, $C_{s} \in H_{w}^{1} (\Omega )$, to problem (\ref{10}) and (\ref{6}), (\ref{8}), (\ref{9}) in the above sense.

Considering the process of pollutant transport and diffusion in the atmosphere (and in the water) the concentrations
$C_{1}, C_{2},\ldots,C_{S}$ of pollutants can not be negative if they are non-negative in the initial state $t=0$ for
all $\xi \in (0,1)$. This property is called
{\it{non-negativity preservation}} and it is well studied for single heat-diffusion equation.
Unfortunately, one needs additional assumptions that the system has a (quasi-) monotonicity property, also called
{\it{cooperativeness}}. This condition is rather restrictive and in this paper we use an idea from
{\cite {13}} to establish maximum principle for the system (\ref{10}).

\noindent
{\bf{Theorem 2.}} Let $C$ be a solution of the system (\ref{10}). Assume that for all $s=1,2, \dots,S$:
a) $ f_{s} ( t, \xi ) \geq 0, \; (t,\xi) \in Q_T$; b)  $B_{s} (\xi ,C)$ is Lipshitz continuous with respect to the concentrations $C$ and it satisfies the inequality
$B_{s} ( \xi, C_1,\dots,C_{s-1},0,C_{s+1},\dots,C_S) \geq 0$,
$\forall \; C \in R_{+}^{S} \equiv
\{ C_{s} \geq 0 \}$. Then $C_{s} \geq 0$ for all $t \in [0,T]$ and $\xi \in [0,1]$.
If we further assume that c) there exists $M_{s} \geq 0$ such that
$B_{s} ( \xi, C_1,\dots,C_{s-1},M_s,C_{s+1},\dots,C_S) \leq 0,\; \forall \; C \in R_{+}^{S}$,
then $0 \leq C_{s} ( t , \xi ) \leq M_{s} < \infty$,  $s =1, \dots, S$.

\section{Fitted finite volume difference scheme for the transformed problem}

Let the interval $[0,1]$ be subdivided into  $N$ intervals $I_i = [\xi _i ,\xi _{i + 1}]$, $i = 1,2,\ldots ,N$ with $0 = \xi _1 < \xi _2 < \ldots < \xi _N < \xi _{N + 1} = 1$ and
$h_i = \xi _{i + 1} - \xi _i $. We  set
$\xi _{i - \frac{1}{2}} = 0.5\left( {\xi _{i - 1} + \xi _i } \right)$,
$\xi_{i + \frac{1}{2}} = 0.5\left( {\xi _i + \xi _{i + 1} } \right)$,
$\hbar _i = \xi _{i + \frac{1}{2}} - \xi _{i - \frac{1}{2}} $ for $i =2,3,\ldots,N$.

{\it{A. Internal nodes}.}
We integrate equation (\ref{10}) on the cell $[\xi _{i - \frac{1}{2}} ,\xi _{i + \frac{1}{2}}]$:
\begin{multline}\label{11}
\int\limits_{\xi _{i -\frac{1}{2}} }^{\xi _{i + \frac{1}{2}} } {\frac{\partial
C_s}{\partial t}d\xi = \int\limits_{\xi _{i - \frac{1}{2}} }^{\xi _{i + \frac{1}{2}} }
{\frac{\partial }{\partial \xi }\left( {p_s(\xi )\frac{\partial C_s
}{\partial \xi } + q_s(\xi )C_s } \right)d\xi } } \\
+ \int\limits_{\xi _{i - \frac{1}{2}
} }^{\xi _{i + \frac{1}{2}} } {\left[ {B_s(\xi,C_1,\ldots,C_S) + f_s(t,\xi)} \right]d\xi}.
\end{multline}
Applying the mid-point quadrature rule to all the integrals in (\ref{11}) with
exception to the second one we obtain
\begin{multline}\label{12}
\left. {\frac{\partial C_s }{\partial t}} \right|_{(t, \xi _i)} \hbar _i
= \left. {\left( {p_s(\xi )\frac{\partial C_s }{\partial \xi } + q_s(\xi )C_s
} \right)} \right|_{\left( {t, \xi _{i + \frac{1}{2}}} \right)}\\
 - \left. {\left({p_s(\xi )\frac{\partial C_s}{\partial \xi } + q_s(\xi )C_s } \right)}
\right|_{\left(t, {\xi _{i - \frac{1}{2}}} \right)}
 +\hbar _i \left. {\left[ {B_s(\xi,C_1,\ldots,C_S) + f_s(t,\xi)} \right] } \right|_{(t,\xi _i)}.
\end{multline}
Further, at the derivation of the discrete equations we follow the methodology in {\cite{7}}.
Let us rewrite the equation (\ref{12}) in the form
\begin{equation}\label{13}
\frac{\partial C_{s,i }}{\partial t}\hbar _i = (1 - \xi _{i + \frac{1}{2}
}^2 )\rho_{s,i + \frac{1}{2}} - (1 - \xi _{i - \frac{1}{2}}^2 )\rho_{s,i - \frac{1}{2}} +
\hbar_i \left( {B_{s,i} + f_{s,i}} \right),
\end{equation}
where
\begin{equation}\label{14}
\rho_s \equiv \rho_s \left( C_s \right) = a^2(1 - \xi ^2)k_s(\xi )\frac{\partial C_s}
{\partial \xi } + (2a^2\xi k_s(\xi ) - aw )C_s.
\end{equation}
We need to derive an approximation of the continuous flux $\rho_s$ in the point
$\xi _{i + 1 / 2} $, $i =2,3,\ldots ,N - 1$.
To do this, we consider the {\it{two-point BVP}}
\begin{gather}
\left( {l_{s,i + \frac{1}{2}} (1 - \xi
^2){V_s}' + m_{s,i + \frac{1}{2}} V_s} \right)^\prime = 0,\quad \xi \in I_i,\label{15}\\
V_s(\xi _i ) = C_{s,i} ,\quad \quad V_s(\xi _{i + 1} ) =C_{s,i+1},\label{16}
\end{gather}
where $l_s = a^2k_s(\xi )$, $m_s = 2a^2 \xi k_s(\xi ) - a$, $l_{s,i + \frac{1}{2}} = l_s(\xi _{i + \frac{1}{2}} )$,
$m_{s,i + \frac{1}{2}} = m_s(\xi _{i + \frac{1}{2}} )$.
Integrating (\ref{15}) yields the first-order linear equation.
Solving this equation and using the boundary conditions (\ref{16}) gives
\begin{equation}\label{17}
\rho_{s,i + \frac{1}{2}}=m_{s,i + \frac{1}{2}} \frac{\Delta_{s,i} \left(
{\xi _{i + 1} } \right)C_{s,i + 1} - \Delta_{s,i} \left( {\xi _i }
\right)C_{s,i} }{\Delta_{s,i} \left( {\xi _{i + 1} } \right) - \Delta_{s,i} \left( {\xi _i } \right)},
\end{equation}
where
$\alpha _{s,i} = \frac{m_{s,i + \frac{1}{2}} }{l_{s,i + \frac{1}{2}} }$, 
$\Delta_{s,i} \left( {\xi _i } \right) =
\left( {\frac{1 + \xi _i }{1 - \xi _i }} \right)^{\frac{\alpha _{s,i}}{2}}$.
In a similar way we approximate $\rho_{s,i - \frac{1}{2}}$ in (\ref{13}) for $2 \le i
\le N$.
For approximation of $\rho_{s, N+\frac{1}{2}}$ we solve the BVP
\begin{gather*}
\left( {\bar {l}_{s,N + \frac{1}{2}} (1 - \xi ){V_s}' + m_{s,N + \frac{1}{2}} V_s}
\right)^\prime = M_2, \quad
V_s(\xi_N ) = C_{s,N},\quad V_s(\xi _{N + 1} ) =0, 
\end{gather*}
where $\bar {l}=a^2(1+\xi)k_s(\xi)$. After some calculation for the flow
$\rho_{s,N + \frac{1}{2}}$ we get
\[
\rho _{s,N +\frac{1}{2}}
= 0.5\left[{C_{s,N + 1} \left( {\bar {l}_{s,N + \frac{1}{2}} + m_{s,N + \frac{1}{2}} } \right) -
C_{s,N} \left( {\bar {l}_{s,N + \frac{1}{2}} - m_{s,N + \frac{1}{2}} } \right)} \right].
\]
{\it{B. Boundary nodes.}} To approximate the boundary condition on the left vertical boundary $\xi = 0$
we proceed as for the internal grid nodes, but
integrating  equation (\ref{10}) on the interval $[\xi_1 ,\xi_{\frac{3}{2}} ]$ (i.~e. in the semi-interval by $\xi$) to get
\[
\frac{\partial C_{s,1} }{\partial t}\frac{h_1 }{2} = (1 - \xi _{\frac{3}{2}}^2 )\rho_{s,\frac{3}{2}} - (1 - \xi _1^2 )
\rho_{s,1} + \frac{h_1 }{2}\left( {B_{s,1}+ f_{s,1}} \right).
\]
>From (\ref{17}) for $i=1$ we get the approximation for $\rho_{s,\frac{3}{2}}$.
For $(1 - \xi _1^2 )\rho_{s,1}$, where $\xi_1 = 0$, using
the expression for $\rho_s$ (\ref{14}) and the boundary condition
(\ref{6}) we find
\[
(1 - \xi_1^2 )\rho_{s,1} = a(\delta_s k_s(\xi _1 ) - w )C_{s,1}.
\]
On the right vertical boundary $\xi = 1$ we have $C_{s,N + 1}  = 0.$

Finally, for the space approximation we obtain the ODE non-linear system of equations for $C_{s,i}(t)$,
$s=1,2,\ldots,S$, $i=1,2,\ldots,N+1$:
\begin{gather*}
\frac{\partial C_{s,1} }{\partial t}\frac{h_1 }{2} = - e_{s,1,1} C_{s,1} +
e_{s,1,2} C_{s,2}+\frac{h_1 }{2}\left[B_s(\xi_1,C_{1,1},C_{2,1},\ldots,C_{S,1})+f_{s,1}(t)\right],\\
\frac{\partial C_{s,i} }{\partial t}\hbar_i = e_{s,i,i - 1} C_{s,i - 1} -
e_{s,i,i}C_{s,i} + e_{s,i,i + 1} C_{s,i + 1}\\
+\hbar_i \left[ B_s(\xi_i,C_{1,i},C_{2,i},\ldots,C_{S,i})+f_{s,i}(t)\right],\quad i = 2,3,\ldots ,N , \\
C_{s,N + 1} =0,
\end{gather*}
where
\begin{gather*}
e_{s,1,1} = \frac{(1 - \xi _{\frac{3}{2}}^2 )m_{s,\frac{3}{2}} \Delta _{s,1} (\xi _1
)}{\Delta _{s,1} (\xi_2 ) - \Delta_{s,1} (\xi_1 )} + a\left( {\delta _s
k_s (\xi _1 ) - w} \right);\\
e_{s,i,i - 1} = \frac{(1 - \xi _{i - \frac{1}{2}}^2 )m_{s,i - \frac{1}{2}} \Delta _{s,i
- 1} (\xi _{i - 1} )}{\Delta _{s,i - 1} (\xi _i ) - \Delta _{s,i - 1} (\xi
_{i - 1} )},\; i = 2,3,\ldots ,N;\\
e_{s,i,i + 1} = \frac{(1 - \xi _{i + \frac{1}{2}}^2 )m_{s,i + \frac{1}{2}} \Delta _{s,i}
(\xi _{i + 1} )}{\Delta _{s,i} (\xi _{i + 1} ) - \Delta _{s,i} (\xi _i )},\;
i = 1,2,\ldots ,N - 1;\\
e_{s,i,i} = \frac{(1 - \xi _{i + \frac{1}{2}}^2 )m_{s,i + \frac{1}{2}} \Delta _{s,i}
(\xi _i )}{\Delta _{s,i} (\xi _{i + 1} ) - \Delta _{s,i} (\xi _i )} +
\frac{(1 - \xi _{i - \frac{1}{2}}^2 )m_{s,i - \frac{1}{2}} \Delta _{s,i - 1} (\xi _i
)}{\Delta _{s,i - 1} (\xi _i ) - \Delta _{s,i - 1} (\xi _{i - 1} )},\\
i= 2,3,\ldots ,N - 1; \; e_{s,N,N + 1} = 0.5(1 - \xi _{N + \frac{1}{2}}^2 )
\left( {\bar {l}_{s,N +\frac{1}{2}} + m_{s,N + \frac{1}{2}} } \right),\\
e_{s,N,N} = 0.5(1 - \xi _{N + \frac{1}{2}}^2 )\left( {\bar {l}_{s,N +\frac{1}{2}}
- m_{s,N + \frac{1}{2}} } \right) + \frac{(1 - \xi _{N - \frac{1}{2}}^2 )m_{s,N -\frac{1}{2}}
\Delta _{s,N - 1} (\xi _N )}{\Delta _{s,N - 1} (\xi _N ) - \Delta _{s,N -1} (\xi _{N - 1} )}.
\end{gather*}
In order to discretize the problem with respect to $t$ we introduce the mesh
\[
0 = t_1 < t_2 < \ldots < t_j < t_{j + 1} < \ldots < t_{M + 1} = T,\quad
\Delta t_j = t_{j + 1} - t_j .
\]
Then, the fully implicit scheme can be written in the form
\begin{gather}
\frac{C_{s,1}^{j + 1} - C_{s,1}^j }{\Delta t_j }\frac{h_1 }{2}
 = - e_{s,1,1} C_{s,1}^{j + 1} + e_{s,1,2} C_{s,2}^{j + 1} \nonumber\\
 + \frac{h_1}{2} \left[ {B_s (\xi _1 ,C_{1,1}^{j + 1},\ldots
,C_{S,1}^{j + 1} ) + f_{s,1}^{j + 1} } \right],\label{18} \\
\frac{C_{s,i}^{j + 1} - C_{s,i}^j }{\Delta t_j }\hbar _i
= e_{s,i,i - 1} C_{s,i - 1}^{j + 1} - e_{s,i,i} C_{s,i}^{j + 1} + e_{s,i,i +1} C_{s,i + 1}^{j + 1} \nonumber \\
 +\hbar _i \left[ {B_s (\xi _i ,C_{1,i}^{j + 1},\ldots ,C_{S,i}^{j + 1} ) + f_{s,i}^{j + 1} } \right], \quad i = 2,3,\ldots ,N, \label{19}\\
 C_{s,N + 1}^{j + 1} = 0,\quad s=1,2,\ldots,S, \label{20}
 \end{gather}
 where $C_{s,i}^{j}=C_s(t_j,\xi_i)$. To solve the non-linear system (\ref{18})--(\ref{20}) we have used
 Newton's method, which leads to a linear system of equations.

\section {Numerical Experiments}

To show the efficiency and usefulness of the discretization method, various test problems with
different choices of parameters were solved. In the numerical experiment we approximate the Dirac-delta function by the function
\begin{gather*}
d_h (\xi ) = \left\{ {{\begin{array}{*{20}c}
 {\frac{2h - \left| {\xi^\ast - \xi } \right|}{4h^2},\quad \xi \in \left(
{\xi^\ast - 2h,\xi ^\ast + 2h} \right),} \hfill \\
 {0,\quad \quad \quad \quad \; \; \xi \notin \left( {\xi ^\ast - 2h,\xi
^\ast + 2h} \right).} \hfill
\end{array} }} \right.
\end{gather*}
\begin{figure}[htbp]
\hfill%
\begin{minipage}[b]{0.5\textwidth}
\centering
\includegraphics[width=\textwidth]{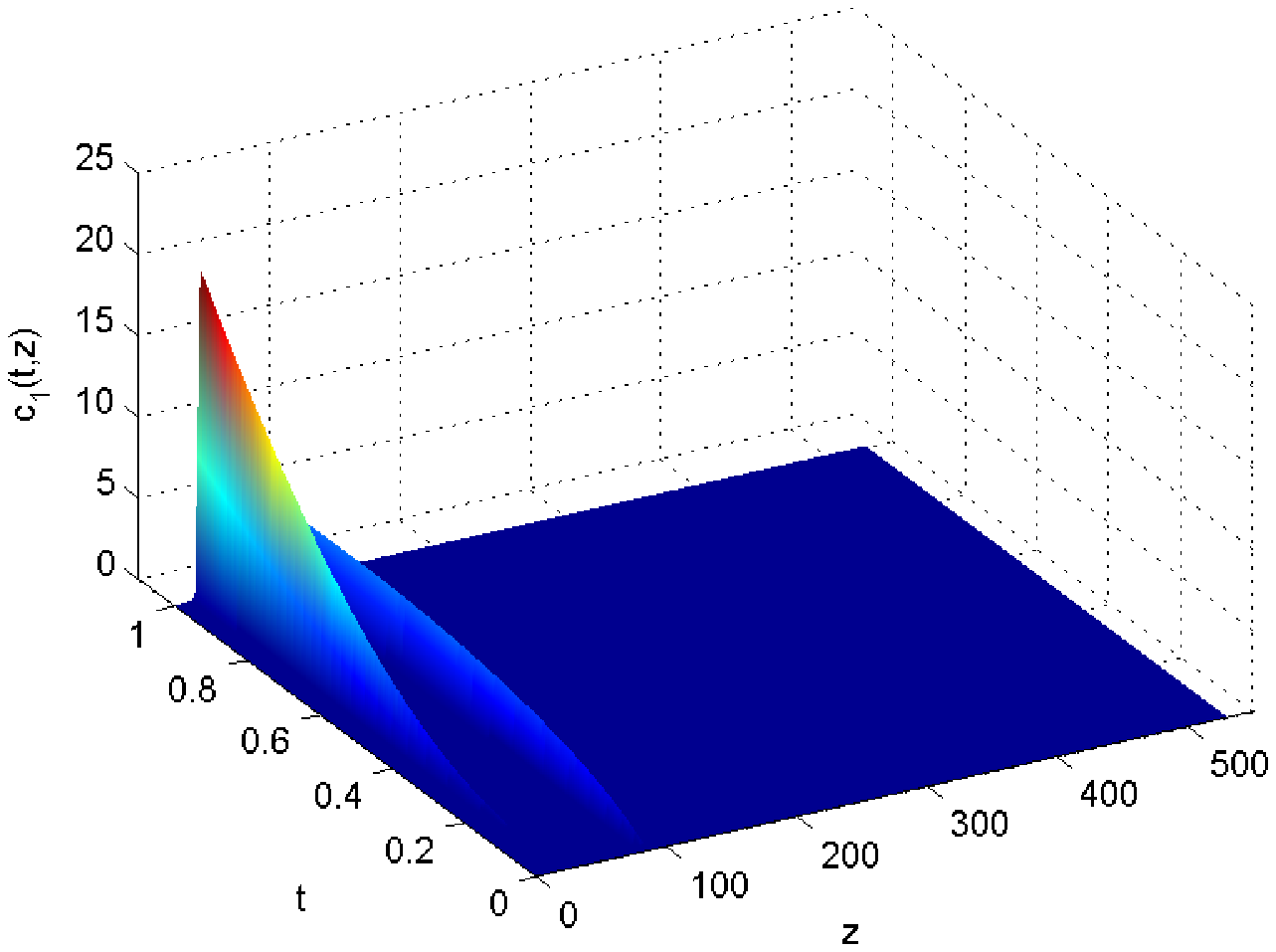}
\caption{Numerical solution $c_1(t,z)$.}\label{f1}
\end{minipage}%
\hfill%
\begin{minipage}[b]{0.5\textwidth}
\centering
\includegraphics[width=\textwidth]{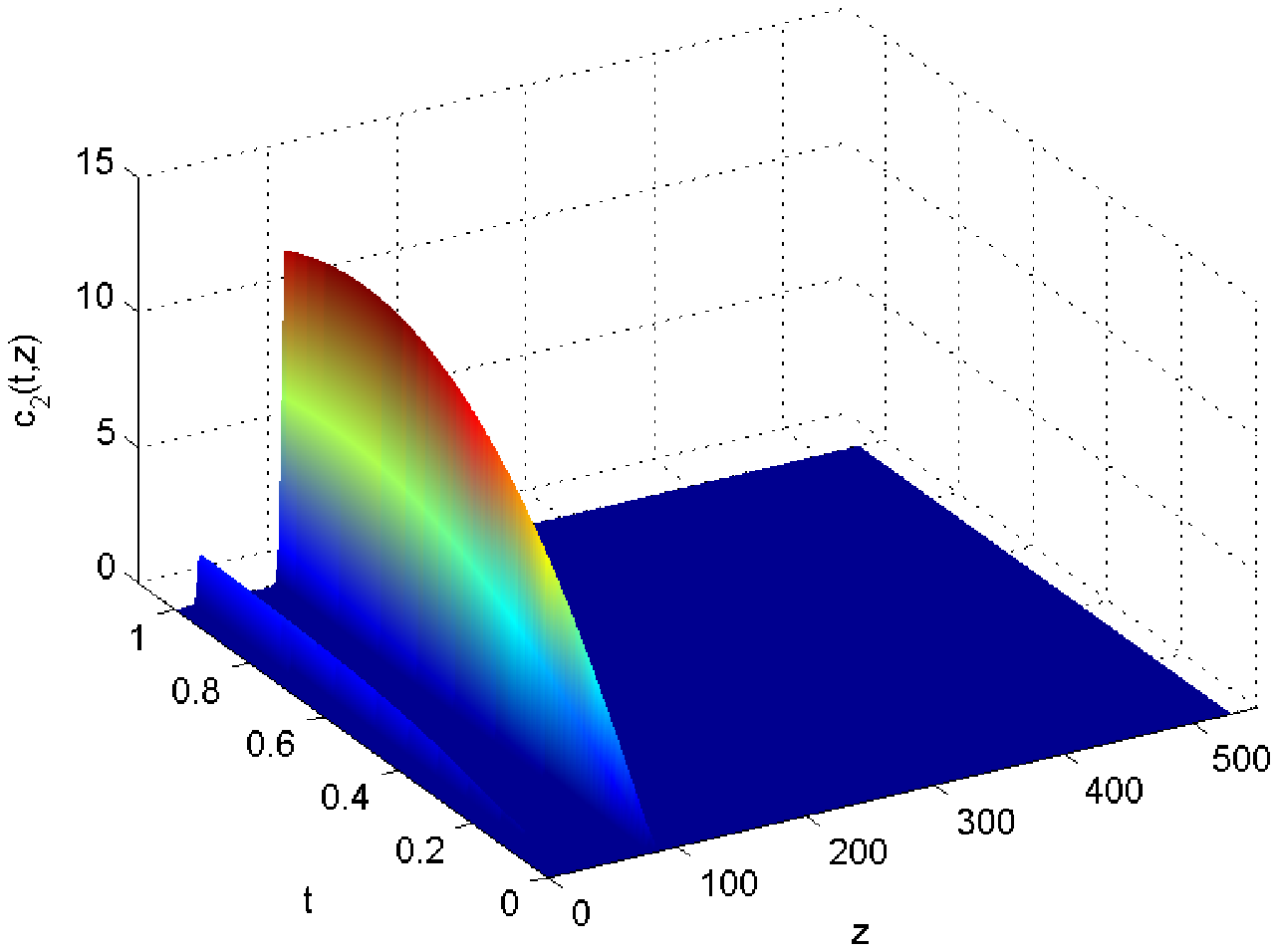}
\caption{Numerical solution $c_2(t,z)$.}\label{f2}
\end{minipage}\hfill\hbox{}%
\end{figure}
\begin{figure}[htbp]
\begin{center}
\includegraphics[width=0.5\textwidth]{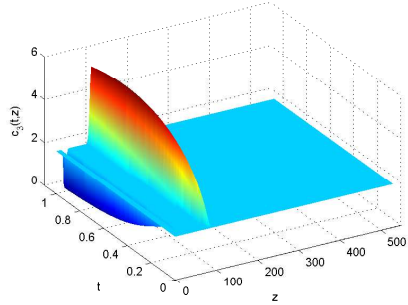}
\caption{Numerical solution $c_3(t,z)$.}\label{f3}
\end{center}
\end{figure}
For the numerical results presented here, we have used  the following functions and values of the coefficients in
the problem under consideration:
$S=3$, $K_1(z)=1$, $K_2(z)=K_3(z)=5$, $w=1$, $Q_1(t)=t$, $Q_2(t)=1-t$, $Q_3(t)=0$, $z_1^\ast=20$, $z_2^\ast=85$,
$\delta_1=\delta_2=\delta_3=0$, $T=1$, $c_{1,0}=c_{2,0}=0$, $c_{3,0}=2$, $a=0.005$, $R_s=\gamma_{s,2}c_2+\beta_{s,1,3}c_1c_3$, $s=1,2,3$, where
$\gamma_{1,2}=-\gamma_{2,2}=\gamma_{3,2}=2000$, $\beta_{1,1,3}=-\beta_{2,1,3}=\beta_{3,1,3}=-1000$.
A part of these data are taken from {\cite{8}}. It is easy to check that the conditions a), b) and c) of Theorem 2 are fulfilled. Fig. \ref{f1}, \ref{f2} and \ref{f3} show the numerical computed concentrations $c_1(t,z)$, $c_2(t,z)$ and $c_3(t,z)$.

We have used the Runge method for practical estimation of the rate of convergence
of the scheme with {\it{respect to the space variable}} at fixed value of $t=T=1$.
We have used three inserted grids with 100, 200 and 400 subintervals respectively by $\xi$ and $\Delta t=\Delta t_j =0.001$. A part of the results from the calculations for the rate of convergence are presented in Table 1.
\begin{table}[htbp]
\centering
\caption{ Runge method for the rate of convergence $n_s$, $s=1,2,3$.}
\begin{tabular}
{|p{10pt}|p{21pt}|p{21pt}|p{21pt}|p{22pt}|p{21pt}|p{17pt}|p{17pt}|p{17pt}|p{17pt}|p{17pt}|p{17pt}|p{17pt}|p{20pt}|p{17pt}|p{17pt}|}
\hline
$\xi$&
0.00&
0.01&
0.02&
0.03&
0.04&
0.05&
0.06&
0.07&
0.08&
0.09&
0.10&
0.11&
0.12&
0.13&
0.14 \\
\hline
$n_1$&
11.04&
9.27&
7.63&
6.16&
4.89&
3.84&
3.04&
2.49&
2.26&
3.08&
0.08&
1.40&
2.12&
2.27&
2.71 \\
\hline
$n_2$&
11.04&
9.28&
7.63&
6.16&
4.89&
3.84&
3.04&
2.46&
2.07&
1.40&
1.25&
0.54&
1.57&
2.13&
2.69 \\
\hline
$n_3$&
1.74&
1.07&
2.17&
-1.67&
5.15&
3.84&
3.04&
2.46&
2.07&
1.40&
1.25&
0.54&
1.57&
2.13&
2.69 \\
\hline
\hline
$\xi$&
0.26&
0.27&
0.28&
0.29&
0.30&
0.35&
0.36&
0.37&
0.38&
0.39&
0.40&
0.41&
0.42&
0.43&
0.44 \\
\hline
$n_1$&
23.97&
19.62&
17.23&
14.96&
12.79&
4.00&
2.90&
2.15&
1.89&
1.99&
1.67&
1.19&
1.59&
1.69&
2.36 \\
\hline
$n_2$&
23.98&
19.63&
17.24&
14.97&
12.80&
4.01&
2.93&
2.24&
1.97&
2.03&
1.79&
1.40&
1.68&
1.79&
2.44 \\
\hline
$n_3$&
2.01&
2.00&
1.94&
0.44&
5.25&
2.57&
2.06&
1.79&
1.82&
1.48&
2.79&
2.17&
-1.96&
1.64&
1.71 \\
\hline
\end{tabular}
\end{table}

\section{ Conclusions }

In this work we have considered a one-dimensional nonlinear problem of air pollution. We have used a log-transformation that makes the original problem, defined on a semi-infinite interval, to another one on the interval $(0,1)$. We have discussed the well-posedness of the transformed problem and the non-negativity of its solution. We have derived  a fitted volume difference scheme that preserves the non-negativity property of the differential problem solution as numerical experiments show. Iterative algorithms and two-grid techniques for first solution of the non-linear system \eqref{18}-\eqref{20} , similar to those in \cite{K12,K14,K15} will be used in the future research. Detail experimental and theoretical analysis will be very interesting.

{\bf {Acknowledgements:}} The first author is supported by the Sofia University Foundation
under Grant No 111/2014. The second author is supported by Bulgarian National Fund of Science under
Project DID 02/37-2009.

\end{document}